\documentclass[10pt]{article}
\usepackage[BOE]{express}

\begin{document}

\title{Beyond backscattering: Optical neuroimaging by BRAD}

\author{Pablo Eugui,\authormark{1,*} Antonia Lichtenegger,\authormark{1} Marco Augustin,\authormark{1} Danielle J. Harper,\authormark{1} Martina Muck,\authormark{1,2} Thomas Roetzer,\authormark{2} Andreas Wartak,\authormark{1} Thomas Konegger,\authormark{3} Georg Widhalm,\authormark{4} Christoph K. Hitzenberger,\authormark{1} Adelheid Woehrer\authormark{2} and Bernhard Baumann\authormark{1}}

\address{\authormark{1}Center for Medical Physics and Biomedical Engineering, Medical University of Vienna, Vienna, Austria\\
\authormark{2}Institute of Neurology, General Hospital and Medical University of Vienna, Vienna, Austria\\
\authormark{3}Institute of Chemical Technologies and Analytics, TU Wien, Vienna, Austria\\
\authormark{4}Department of Neurosurgery, General Hospital and Medical University of Vienna, Vienna, Austria}

\email{\authormark{*}pablo.euguiarrizabalaga@meduniwien.ac.at} 



\begin{abstract}
Optical coherence tomography (OCT) is a powerful technology for rapid volumetric imaging in biomedicine. The bright field imaging approach of conventional OCT systems is based on the detection of directly backscattered light, thereby waiving the wealth of information contained in the angular scattering distribution. Here we demonstrate that the unique features of few-mode fibers (FMF) enable simultaneous bright and dark field (BRAD) imaging for OCT. As backscattered light is picked up by the different modes of a FMF depending upon the angular scattering pattern, we obtain access to the directional scattering signatures of different tissues by decoupling illumination and detection paths. We exploit the distinct modal propagation properties of the FMF in concert with the long coherence lengths provided by modern wavelength-swept lasers to achieve multiplexing of the different modal responses into a combined OCT tomogram. We demonstrate BRAD sensing for distinguishing differently sized microparticles and showcase the performance of BRAD-OCT imaging with enhanced contrast for ex vivo tumorous tissue in glioblastoma and neuritic plaques in Alzheimer's disease.
\end{abstract}



\section{Introduction}
Light scattering is fundamental to modern optical imaging approaches in biomedicine. While posing essential ranging limitations to fluorescence based light sheet microscopy (which are now partly overcome by tissue clearing) \cite{Keller2012,Richardson2015}, light scattering is the basis for image contrast in applications ranging from tissue contrasting to perfusion imaging \cite{Tuchin2007}. Optical coherence tomography (OCT) is an imaging technique based on the detection of backscattered light \cite{Huang1991}. Augmented by an interferometric detection principle of low-coherent (i.e. broadband) light \cite{Fercher1986}, OCT provides rapid three-dimensional (3D) imaging with micrometer-scale resolution. Consequently, OCT has found various applications ranging from established routine diagnostics in ophthalmology and cardiology to neuroimaging \cite{Fercher2003, Kut2015, Lichtenegger2017, Badon2016, Alex2015}. OCT is sensitive to subtle differences of the backscattered signal from diverse healthy and pathological tissues and provides sufficient contrast in many imaging applications. Still, state-of-the-art OCT systems have limited capabilities for revealing the nature of light scattered at angles other than that opposing the illumination direction. The illumination and detection unit of most OCT devices is based on a single-mode fiber which serves as a pinhole and essentially integrates all photons scattered by the sample into the numerical aperture (NA) of the fiber. However, light scattering of biological tissues often is subject to a pronounced directionality which strongly depends on their microstructural composition \cite{Jacques2013}. Recently developed dark-field OCT approaches aimed to overcome this limitation of standard OCT \cite{Villiger2010, Blatter2011, Auksorius2015}. Rather than imaging in a bright field configuration, these systems omitted the directly backscattered light component by means of a sophisticated yet quite complex optical layout and only detected light scattered into a specific cone. \par
Few-mode fibers (FMFs) recently revolutionized the field of telecommunications. By multiplexing signals into several fiber modes, data transmission can be parallelized and sped up significantly \cite{Randel2011, Richardson2013}. Step-index FMFs transmit a small number of transverse modes, each of which has characteristic dispersion properties, such that light signals coupled into different modes have dissimilar propagation speeds \cite{Snyder2012}. Every transverse fiber mode also governs a specific transverse field distribution and can therefore be excited by a specific angular intensity pattern. Aside from multiplexed signal transduction in telecommunication applications, the well-defined transmission characteristics of FMFs have opened the door to sensing and imaging approaches such as temperature and strain mapping, advanced transmitted light microscopy, and cellular tracking \cite{Weng2015, Kreysing2014,Cizmar2012}. Here we show that the unique features of FMFs enable simultaneous bright and dark field (BRAD) imaging for OCT. By decoupling illumination and detection paths, we gain access to the directional scattering signatures of different tissues, as backscattered light is picked up by the different modes of a FMF depending upon the angular scattering distribution. We then exploit the distinct modal propagation properties of the FMF in concert with the long coherence lengths provided by modern wavelength-swept lasers in order to achieve multiplexing of the different modal responses into a combined OCT image. We demonstrate BRAD sensing for distinguishing differently sized scattering particles and showcase the performance of BRAD-OCT imaging with enhanced contrast for tumorous tissue in glioblastoma and for neuritic plaques in brain tissue affected by Alzheimer's disease.

\section{Materials and Methods}

\subsection{Few-mode fiber detection concept}
\begin{figure*}[!htb]
\centering
\includegraphics[width=1\linewidth]{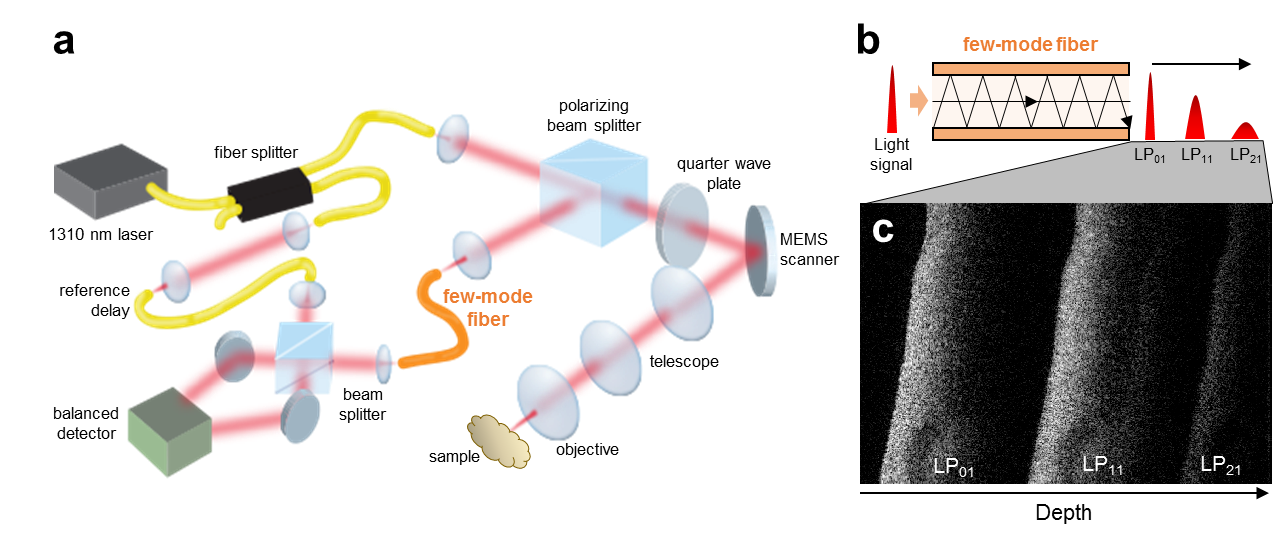}
\caption{FMF detection concept. (a) Simplified scheme of the BRAD-OCT experimental set-up with FMF detection. The FMF implemented in the prototype collects the backscattered light and encodes it into the different transversal modes. As light portions coupled into the modes travel different path lengths along the FMF as shown in (b), they will sequentially arrive at the distal end of the fiber, reconstructing the modal images at different depth positions. This concept is demonstrated in (c) where the modal images are mapped into different depths of a cross-sectional OCT image of brain tissue.}
\label{fig:fig1}
\end{figure*}

Our approach is based on a modification of the interferometric layout conventionally used for OCT (see Fig.  \ref{fig:fig1}(a)). State-of-the-art OCT interferometers are based on single-mode fiber (SMF) optics and essentially split light from a low-coherent source into a sample arm and a reference arm \cite{Klein2017}. The light beam in the sample arm illuminates the sample via dedicated imaging optics and is scattered back by the tissue before being fed back into the fiber-optic interferometer. At the output of the interferometer, the sample beam interferes with the reference beam which was itself passed through a delay line matching the length of the sample beam path. In the case of an OCT system using a wavelength-swept laser as a light source (i.e. swept-source OCT), the interferometer commonly follows a Mach-Zehnder layout. The spectral interference pattern is recorded at the detection unit as a function of time and wavenumber, and depth scan information is retrieved by an inverse Fourier transform of the interference spectrum \cite{Fercher1995}. Swept-source OCT provides imaging at depth scan rates of 100 kHz and beyond with millimeter scale ranging depths, micrometer scale resolution, and high detection sensitivity for scattered light with intensities up to $\sim$ $10^{-10}$ (100 dB) weaker than that of a mirror reflection \cite{Klein2017}. \par
In our approach, we decouple sample illumination and detection paths as shown in Fig. \ref{fig:fig1}(a). While light is transmitted by SMF optics at the source output, reference arm and sample illumination path, light backscattered by the sample is picked up by a step index few-mode fiber \cite{Eugui2017}. FMFs differ from SMFs in that they transmit multiple modes simultaneously. Each mode (i) possesses a specific field pattern $\psi_{lm}$ such that it accepts and emits particular light fields only and (ii) has a particular propagation constant  $\beta_{lm}$ and therefore transmits light coupled into a specific mode at a different speed than all other modes (Fig. \ref{fig:fig1}(b)) \cite{Snyder2012}. As a detection fiber in our setup, the FMF thus performs a modal decomposition of the backscattered field and guides the light components received by the fiber modes via different effective path lengths to the detection unit. Upon interference with the reference beam, sample field components backscattered at distinct directions and launched into different modes appear at different depths in the multiplexed OCT tomogram as shown in Fig. \ref{fig:fig1}(c).

\subsection{Description of few-mode fiber detection for OCT imaging}

\begin{figure*}[!htbp]
\centering
\includegraphics[width=1\linewidth]{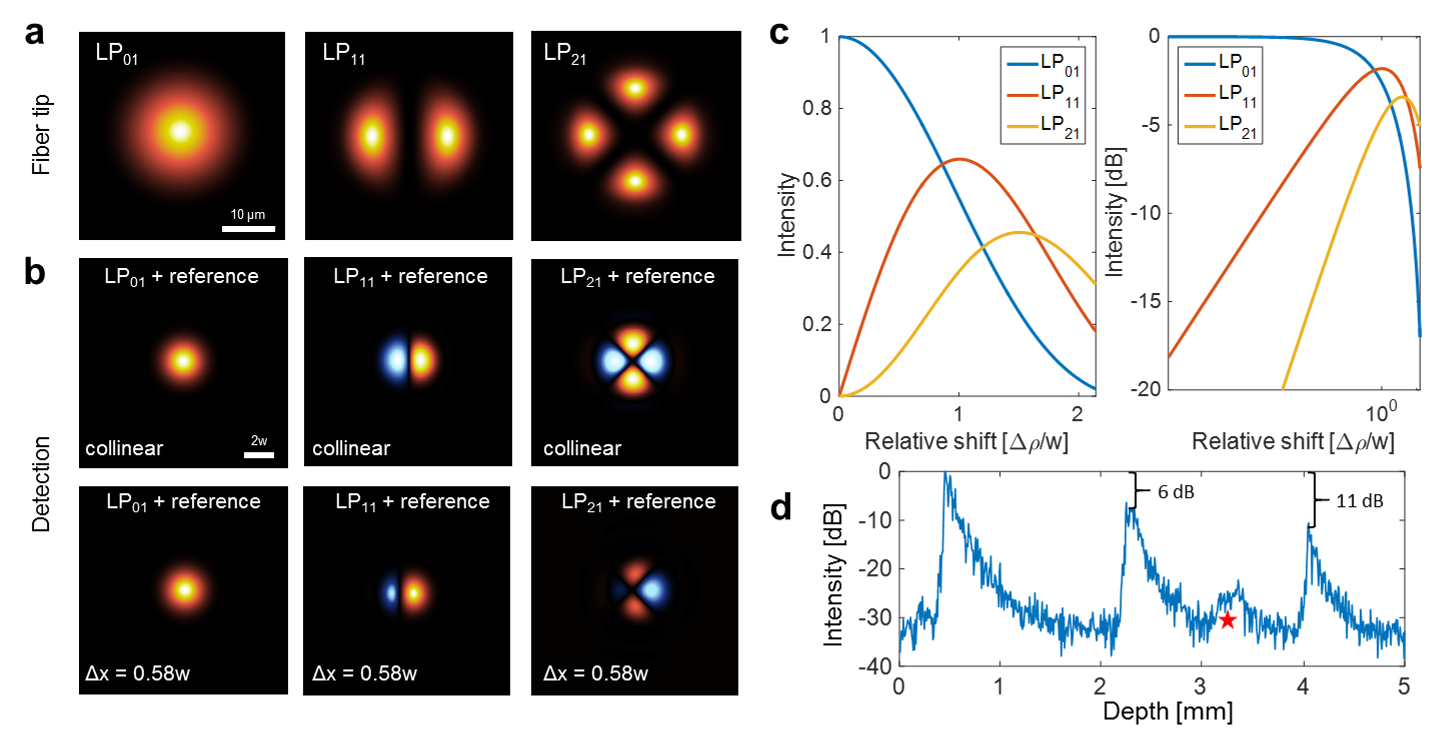}
\caption{Interference of fiber modes.(a) Intensity distributions of modes $LP_{01}$, $LP_{11}$ and $LP_{21}$ supported by the FMF. (b) Interference beam profiles simulated for a perfect alignment (upper row) and for an asymmetric interference pattern where the spots have been displaced by 0.58\textit{w} (bottom row). (c) Modal intensities with respect to the relative shift $\Delta \rho/\textit{w}$  between sample and reference beam. (d) Relative signal-to-noise ratio for the different FMF modes measured with a scattering paper in the sample position. The red star indicates a higher-order mode that appears close to the noise level in our interference signal. This mode was disregarded in the imaging experiments due to its weak intensity and strong dispersion.}
\label{fig:S1}
\end{figure*}

The modes of an optical fiber with a cylindrically symmetrical refractive index profile are given by the solutions of Helmholtz's equation in radial coordinates \cite{Snyder2012}. Step index fibers with core radius \textit{a}, core refractive index $n_{core}$  and cladding refractive index $n_{clad}$ are called weakly guiding in case of a very small profile height parameter, \(\Delta=(n_{core}^2-n_{clad}^2)/2n_{core}^2\ll1\). Since guided rays propagate approximately parallel to the fiber axis (i.e. paraxial) in weakly guiding fibers, the transverse field components are far stronger than the longitudinal ones such that the waves can be described as approximately transverse electromagnetic (TEM) ones. For a fiber oriented in z direction, the linear polarization states in the x and y direction are then orthogonal and form linearly polarized (\textit{l,m})  modes ($LP_{lm}$) \cite{Snyder2012,Teich1991,Pal1992}. The mode fields supported by weakly guiding step index fibers are given by

\begin{equation} \label{eq:1}
     \psi_{l,m}(r,\varphi)=\left\{
                \begin{array}{ll}
                  \frac{A_{lm}}{J_l(U_{lm})}J_l(U_{lm}\frac{r}{a})e^{\pm il \varphi}, \quad \quad r<a\\ 
                 \\ \frac{A_{lm}}{K_l(W_{lm})}K_l(W_{lm}\frac{r}{a})e^{\pm il \varphi}, \quad r<a
                \end{array}
              \right.
\end{equation}

where $J_l$ are the Bessel functions of the first kind and \textit{l}-th order and $K_l$ are the modified Bessel functions of the second kind and \textit{l}-th order. \(U_{lm}=a\sqrt{k^2n^2_{core}- \beta_{lm}^2}\) and   are the core and cladding parameters, $A_{lm}$  denotes the modal amplitude, and \(W_{lm}=a\sqrt{\beta_{lm}^2-k^2n^2_{clad}}\) is the mode-dependent propagation constant \cite{Snyder2012}. Simulated intensity distributions of modes $LP_{01}$, $LP_{11}$ and $LP_{21}$ supported by the FMF used for our experiments are shown in Fig. \ref{fig:S1}(a). The modal content of a fiber is governed by the \textit{V} parameter given by \(V=\sqrt{U^2+W^2}\) \cite{Snyder2012,Pal1992}. Depending on the shape of the field launched into the fiber, supported fiber modes will be excited and contribute to light propagation with different weightings. The field at the fiber input can be described as the superposition of the transmitted mode fields: 

\begin{equation} \label{eq:2} \Psi(r,\varphi,z,t)=\sum_pa_p\psi_p(r,\varphi)e^{i(\omega
t-\beta_pz-2kz_s)}\end{equation}

where $\beta_p$ is the propagation constant of the \textit{p}-th fiber mode with \(kn_{core}\geq \beta_p \geq kn_{clad}\) and 2$kz_s$  is the phase shift related to the sample position. The weights $a_p$ account for the share of the field transmitted by the \textit{p}-th mode. Assuming that leakage into other modes is negligible for a few-mode fiber without significant twists or bends \cite{Snyder2012} and that birefringence effects and (mode-dependent) transmission loss can be neglected for the rather short fiber length used in our setup, the modal fields $\psi_p$  will be transmitted with preserved weights $a_p$. Owing to the mode-dependent propagation constants $\beta_p$ , the mode fields travel different optical path lengths  $\beta_pL_{\text{FMF}}/k$ through the FMF with length  $L_{\text{FMF}}$. At the distal end of the FMF located in the detection unit, light transmitted by the FMF is collimated by a lens which corresponds to a 2D Fourier transform of the transverse field. By denoting the Fourier transform of the \textit{p}-th field component as \(\psi_p(r,\varphi)  \overset{FT}{\longleftrightarrow}\tilde{\psi}_p( \rho,\theta)\) and by using the linearity of the Fourier transform, the collimated field reads 

\begin{equation} \label{eq:3} \tilde{\Psi}(\rho,\theta,z,t)=\sum_pa_p\tilde{\psi}_p(\rho,\theta)e^{i(\omega t-\beta_pL_{\text{FMF}}-2kz_s)}.\end{equation}
The reference light beam is transmitted by a single-mode fiber with matched length $L_{\text{ref}}=L_{\text{FMF}}$ and propagation constant $\beta_{\text{ref}}$.  After collimation, the reference wave is described by

\begin{equation} \label{eq:4} \tilde{\Psi}_{\text{ref}}(\rho,\theta,z,t)=a_{\text{ref}}\tilde{\psi}_{\text{ref}}(\rho,\theta)e^{i(\omega t-\beta_{\text{ref}}L_{\text{FMF}}-2kz_{\text{ref}})}.\end{equation}

The fields from sample and reference arm are superposed at the detection unit's beam splitter, 

\begin{equation} \label{eq:5}
\begin{split}
\tilde{\Psi}_{det}(\rho,\theta,z,t)=\tilde{\Psi}(\rho,\theta,z,t)+\tilde{\Psi}_{\text{ref}}(\rho,\theta,z,t)\\
=\sum_pa_p\tilde{\psi}(\rho,\theta)e^{i(\omega t-\beta_pL_{\text{FMF}}-2kz_s)}\\
+a_{\text{ref}}\tilde{\psi}_{\text{ref}}(\rho,\theta)e^{i(\omega t-\beta_{\text{ref}}L_{\text{FMF}}-2kz_{\text{ref}})}.
\end{split}\end{equation}

The light intensity of the emerging beam is proportional to the time-average of the squared superposed field, \(I_{det} \varpropto \langle \tilde{\Psi}_{det}^2 \rangle\)  \cite{Born2013}. Hence, the following components contribute to the observed light beam intensity pattern:

\begin{equation} \label{eq:6} I_1(\rho,\theta)\sim a_{\text{ref}}^2 \vert \tilde{\psi}_{\text{ref}}(\rho,\theta)\vert^2, \end{equation}
\begin{equation} \label{eq:7} I_2(\rho,\theta)\sim \sum_pa_p^2\vert \tilde{\psi}_{p}(\rho,\theta)\vert^2, \end{equation}
\begin{equation} \label{eq:8} I_3(\rho,\theta)\sim \sum_{p\neq q}2a_pa_q\vert \tilde{\psi}_{p}(\rho,\theta)\Vert \tilde{\psi}_{q}(\rho,\theta)\Vert \gamma_{p,q}(\tau)\vert cos[(\beta_q-\beta_p)L_{\text{FMF}}-\delta_{p,q}], 
\end{equation}
\begin{equation} \label{eq:9} I_4(\rho,\theta,k)\sim \sum_{p}2a_pa_{\text{ref}}\vert \tilde{\psi}_{p}(\rho,\theta)\Vert \tilde{\psi}_{\text{ref}}(\rho,\theta)\Vert \gamma_{p,\text{ref}}(\tau)\vert cos[(\beta_{\text{ref}}-\beta_p)L_{\text{FMF}}+2k\Delta z-\delta_{p,\text{ref}}]. 
\end{equation}

The first component, Eq. (\ref{eq:6}), corresponds to the intensity of the reference beam. The second component, Eq. (\ref{eq:7}), is the sum of the light intensities launched into all fiber modes. The third term, Eq. (\ref{eq:8}), denotes intermodal interference phenomena whose intensity will, in general, depend on the amount of light coupled into the respective modes (i.e., $a_p$  and $a_q$ ), their mutual degree of coherence $\gamma_{p,q}(\tau)$  with  $\tau=(n_{\text{eff}}^{(q)}-n_{\text{eff}}^{(p)})L_{\text{FMF}}/c$ \cite{Born2013}, their phase delay $(\beta_q-\beta_p)L_{\text{FMF}}-\delta_{p,q}$  and thus on wavenumber \textit{k}. The total intensity of these intermodal components should, theoretically, be nil owing to the orthogonality of the fiber modes in an ideal optical fiber \cite{Snyder2012}, 

\begin{equation} \label{eq:10}\int_A\psi_j\psi_kdA=C\delta_{jk}\end{equation}

and likewise,

\begin{equation} \label{eq:11} \int_A\psi_j\psi_k^ \ast dA=C\delta_{jk}\end{equation} 

where \textit{C} is a constant and $\delta_{jk}$  denotes Kronecker's delta. Facilitated for instance by slight fiber bends, intermodal interference has been observed, investigated and exploited in real-world fibers \cite{Turek2000,Spajer1986,Kacik2004}. Still, the intermodal interference signals described in Eq. (\ref{eq:8}) will only weakly contribute as a fixed pattern signal to the OCT image background (which is removed by basic OCT image processing as described in Methods, if these signals exceed the noise level at all). 
	The last component of the interference signal, Eq. (\ref{eq:9}), finally describes the intensity pattern observed across the beam for the interference of light transmitted by the modes of the FMF and light propagated through the single-mode fiber in the reference arm. This signal provides the basis for the FMF-OCT image. The mutual interference signal of the light beams from sample and reference arm depends on the phase delay  \((\beta_{\text{ref}}-\beta_{p})L_{\text{FMF}}+2k\Delta z\), the mutual degree of coherence \(\gamma_{p,\text{ref}}(\tau)\), the propagation properties of the two fibers and the path length difference between sample and reference arm, \(\Delta z=z_{\text{ref}}-z_s\).  The interference spectra described by Eq. (\ref{eq:9}) are subject to a sinusoidal modulation along \textit{k} proportional to the path length difference $\Delta z$. The phase terms \((\beta_{\text{ref}}-\beta_{p})L_{\text{FMF}}=(n_{\text{eff}}^{\text{(ref)}}-n_{\text{eff}}^{(p)})kL_{\text{FMF}}\) add a mode dependent phase ramp with a fixed slope for the \textit{p}-th mode. When depth scans are computed from these modulated interference spectra by means of an inverse Fourier transform along \textit{k}, the modulation frequency bands covered by the modal interference signals will map into different depth ranges of the final OCT image as shown in Fig. 1(c).
	The FMF-OCT signals described in Eq. (\ref{eq:9}) depend on the interference of the reference field $\tilde{\psi}_{ref}(\rho,\omega)$ and the respective field $\tilde{\psi}_p(\rho,\omega)$  of the \textit{p}-th mode, which in most cases have inherently dissimilar field profiles. The single-mode fiber in the reference arm emits a profile similar to that of the $LP_{01}$ mode of the FMF (only with a different beam waist) and effectively interferes with light launched into and emerging from the fundamental mode of the FMF. Compared to the beam emitted by the single-mode fiber, higher-order FMF modes have rather dissimilar field profiles which may interfere with the mode field of the single-mode reference beam as shown in the top row patterns of Fig. \ref{fig:S1}(b). The integrated intensity of the interference signal however would be zero for perfectly aligned (i.e. collinear) sample and reference beams, again due to the orthogonality of fiber modes as described by Eq. (\ref{eq:11}). 
This extinction of higher-order signals can be obviated by generating asymmetric interference patterns by slightly tilting or offsetting the sample and reference beams with respect to each other in the detection unit. The effect of intentionally displacing the spots by a fraction of the beam waist \textit{w} is demonstrated in Fig. \ref{fig:S1}(b) (bottom row). We simulated the impact of relative displacements of sample and reference beam for modes $LP_{01}$, $LP_{11}$ and $LP_{21}$ interfering with an $LP_{01}$ style reference beam (see Fig. \ref{fig:S1}(c)). While for greater displacements $\Delta \rho$  the integrated interference signal of the fundamental $LP_{01}$ mode diminishes, the net interference signals produced by the higher-order $LP_{11}$ and $LP_{21}$ modes increase with maxima around \(\Delta \rho \simeq w\)  and \(\Delta \rho \simeq 1.5w\),  respectively. In other words, the intensity of the higher-order OCT signals can be boosted at the cost of somewhat reduced $LP_{01}$ intensity. In our experimental setup, the signal-to-noise ratios (SNRs) of the $LP_{11}$ and $LP_{21}$ mode are 6 dB and 11 dB lower than the SNR of the fundamental $LP_{01}$ mode, as shown in Fig. \ref{fig:S1}(d).
	The interfered light beams in the detection unit are focused onto the two balanced detectors. This focusing step corresponds to an inverse 2D Fourier transformation of the transverse field. However, while focusing will change the beam profile once more, owing again to the linearity of the Fourier transform and Parseval's theorem, the overall intensity across the beam cross section and its distribution among the different terms described in Eqs. (\ref{eq:6}) to (\ref{eq:9}) will be conserved.  
	
\subsection{Few-mode fiber selection}
Different optical fibers were simulated and tested for our 1310 nm source in order to achieve a few-mode behavior. With known core and cladding refractive indices, it is possible to estimate the modal behavior of an optical fiber using the V-number, which is also directly related to the NA of the given fiber:

\begin{equation}\label{eq:22}
V = \frac{2\pi}{\lambda}a \mathrm{NA} = \frac{2\pi}{\lambda}a \sqrt[]{n_{core}^2-n_{cladding}^2}
\end{equation}
where $\lambda$ is the wavelength in vacuum, $a$ is the core radius and $n$ the respective refractive indices \cite{Snyder2012}. The requirement for single-mode behavior of an optical fiber is that the normalized frequency $V<2.4048$, while for large $V$, the number of modes can be approximated by:

\begin{equation}
M \approx \frac{V^2}{2}
\end{equation}
where $M$ is the number of modes that the fiber will transmit \cite{Snyder2012}.\par

This criterion was used to select different fibers that might be appropriate for our setup. A fiber simulation program (RP Fiber Calculator 2015, RP Photonics) was used for calculating the launching efficiency into the fibers and extracting the effective refractive indices $n_{\text{eff}}$ of the modes that the fiber was transmitting. The optical path length difference between consecutive transmitted modes is given by:

\begin{equation}\label{eq:32}
\Delta z = \Delta n_{\text{eff}}  L
\end{equation}
where $\Delta n_{\text{eff}}$ is the effective refractive index difference between modes and $L$ is the length of the fiber. Thus, the intermodal difference can be calculated and the length of the fiber can be set to achieve enough separation to reconstruct the modal images in the OCT tomogram avoiding information overlapping. Due to different path lengths of the transmitted modes, dispersion effects induced by the fiber material were different for each mode and had to be independently corrected in post-processing.\par
The fiber finally selected for our setup was a commercial step-index multimode fiber with 0.10 NA and $25\mu m$ core diameter (FG025LJA  Multimode Fiber. Thorlabs, Inc). The effective refractive indices calculated for the $LP_{01}$, $LP_{11}$ and $LP_{21}$ modes were 1.449281, 1.448662 and 1.447867 at 1310 nm, respectively.

\subsection{BRAD-OCT imaging setup}

\begin{figure*}[!h]
\centering
\includegraphics[width=1\linewidth]{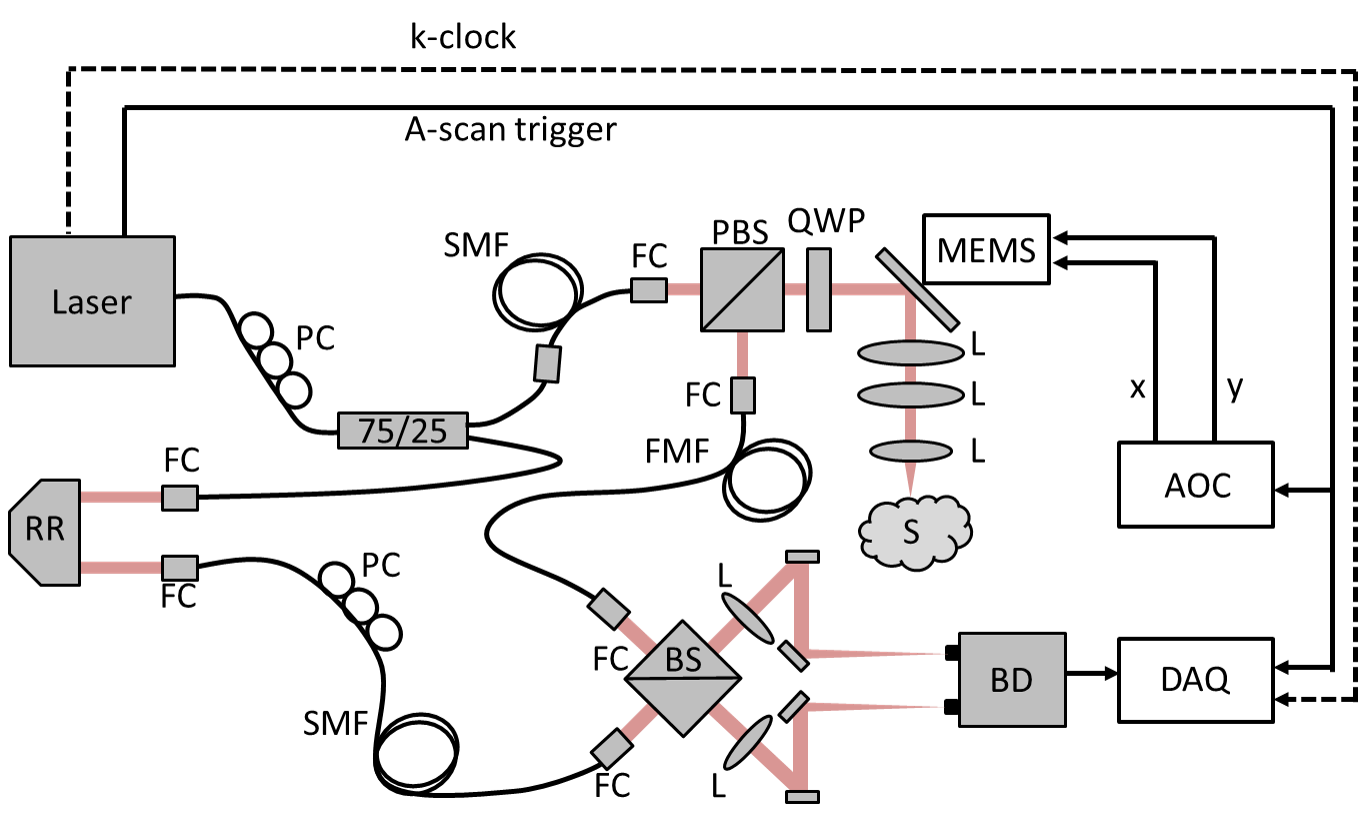}
\caption{Detailed layout of the BRAD-OCT system. RR: retroreflector, PC: polarization control paddles, SMF: single-mode fiber, FMF: few-mode fiber, FC:  fiber collimator, PBS: polarizing beam splitter, BS: beam splitter, QWP: quarter wave plate, L: lens, BD: balanced detector, S: sample, MEMS: microelectromechanical scanning mirror, DAQ: data acquisition board, AOC: analog output card.}
\label{fig:S3}
\end{figure*}

BRAD-OCT was implemented as an experimental setup illustrated in Fig.  \ref{fig:S3}. A short cavity tunable laser (Axsun Technologies, Inc. 1310 nm SS-OCT Laser Engine) with a center wavelength of $\lambda_0$ = 1310 nm and a tuning range of $\Delta\lambda$ = 140 nm was used as the light source. The sweep rate of the laser was 100 kHz with a sampled duty cycle of $\sim$51$\%$. The k-clock signal provided by the light source was used to sample the spectral data at equidistant intervals in k-space. Based on a Mach-Zehnder interferometer layout \cite{Potsaid:10}, the initially linearly polarized laser beam was split by a single-mode fiber coupler, delivering 25\% of the light to a reference arm where a path length delay was introduced to match the length of the sample arm beam path. In the sample arm, light was delivered by a single-mode fiber (Corning SMF-28e+) and launched into a free-space polarization-based circulator consisting of a polarizing beam splitter (PBS) and a quarter wave plate oriented at 45$^\circ$. A polarization control paddle was integrated for the SMF in the sample arm in order to maximize the throughput of horizontally polarized light for the PBS. A micro-opto-electromechanical systems mirror (MEMS; Mirrorcle Technologies, Inc.; diameter: 5.2 mm) was used for 2D scanning and directed the collimated beam through a 50/100 telescope. The beam was focused onto the sample by an achromatic objective lens (f = 50 mm). After double-passing the quarter wave plate, the vertical polarization component of light returning from the sample was reflected by the PBS to be finally coupled into the FMF. This illumination scheme was designed with the aim of maintaining a single-mode Gaussian illumination on the sample while enabling the collection of backscattered light with the FMF. Note that while we do take advantage of the beam's polarization state to efficiently redirect light in the sample arm, our system only detects co-polarized light backscattered by the sample and does not allow us to capture the full polarization state of light as in a polarization sensitive OCT system \cite{DeBoer2017}. \par
Light coupled into the modes of the FMF was interfered with the reference beam in a custom made fiber-connectorized beam splitter cube (OZ Optics, Ltd.), inducing a slight offset between both beams in order to boost the signals of the higher order modes as described in section 2.2. The polarization state of the reference beam was matched to that of the fundamental mode in the FMF using another polarization control paddle for the SMF in the reference arm. We opted for a free-space optics detection because, despite considerable intensity losses produced by the beam splitter and the connectorized collimators, this interference setup was proved to maintain higher-order mode signals of the FMF better than the conventionally used fused fiber splitters \cite{Schmitt1999,Potsaid:10}, which essentially filter out these signals. The two output beams of the beam splitter were guided in free-space by a set of mirrors and focused onto the sensors of a dual balanced photodetector with an active sensor area of 0.15 mm diameter (Thorlabs, Inc., PDB435C) by two achromatic lenses (f = 25 mm). The spectral interference signals were detected with a 350 MHz bandwidth in the balanced photodetector and digitized using the k-clock signal of the laser with a maximum frequency of 350 MHz at 12-bit resolution by a high-speed data acquisition board with a maximum sampling rate of 1.8 GS/s (Alazar Technologies Inc., ATS9360).\par
Polarization in the reference arm was set to maximize the sensitivity of the fundamental mode, inducing a polarization mismatch in the higher modes. The unmatched polarization mode dispersion consequently affected the axial resolution of the higher modes \cite{Jiao2008}, such that the axial resolutions of the system were $\sim$6.5 $\mu$m, 12.0 $\mu$m and 8.0 $\mu$m in air for $LP_{01}$, $LP_{11}$ and $LP_{21}$ respectively, corresponding to $\sim$4.8 $\mu$m, 9.0 $\mu$m and 6.0 $\mu$m in brain tissue assuming a group refraction index of 1.36 \cite{Binding2011}. A lateral resolution of $\sim$8.7 $\mu$m was determined using a US Air Force resolution test target. A sensitivity of 91 dB was measured for the fundamental mode ($LP_{01}$), with a decay of 0.49 dB/mm along a total imaging range of 5.0 mm. The low sensitivity might be caused by the use of non-optimized optics for the few-mode fiber, such as the interference splitter cube where power loss was 3 dB compared to a single-mode fiber for which it was designed. Extra losses were also caused by the FMF itself, where the delivered power is distributed between the different fiber modes unlike a single-mode fiber where all the power is delivered by a unique mode. By imaging a highly scattering sample (Fig. \ref{fig:S1}(d)), sensitivity penalties of -6 dB and -11 dB were estimated for the $LP_{11}$ and $LP_{21}$ modes, respectively. 

\subsection{Acquisition and analysis software}
The data acquisition was performed using custom-made LabView code (LabView 2015, 64-bit, National Instruments). Dedicated programs were implemented for MEMS scanner control, synchronization, data acquisition and multi-plane image preview display. The raw OCT volume scans presented in this work consisted of 1536  $\times$ 512  $\times$ 512 voxels (depth pixels  $\times$ A-scans  $\times$ B-scans) and spanned up to 5.0 mm  $\times$ 5.0 mm  $\times$ 5.0 mm in air.\par
The acquired data was post-processed in Matlab (MATLAB, R2015b, MathWorks). In the first step, standard Fourier domain OCT processing including correction for acquisition trigger fluctuation \cite{Liu2015}, background removal of the raw data sets, spectral shaping \cite{Tripathi2002}, and inverse Fourier transformation was performed \cite{Fercher1995}. Then, an intensity threshold based segmentation algorithm was used for segmenting the sample surface in OCT B-scan images reconstructed by the fundamental FMF mode ($LP_{01}$). As the axial separation of the modal images is given by the effective refractive index $n_{\text{eff}}$ of each mode in combination with the fiber length $L_{\text{FMF}}$ (i.e. it is proportional to ($n^{(1)}_{\text{eff}}$-$n^{(2)}_{\text{eff}}$)$L_{\text{FMF}}$, see Eq. \ref{eq:32}), it remains constant for all the measurements. Thus, the OCT signals of the higher-order $LP_{11}$ and $LP_{21}$ modes were separated by axial displacements of 1.6 mm and 3.3 mm (in air) from the segmented signal of the fundamental mode. Every multiplexed OCT image was cropped to a set of three modal OCT images covering identical depth ranges. \par
Also the higher-order dispersion characteristics are different for every FMF mode. Therefore, once the modal OCT images were segmented and separated, each mode was dispersion compensated independently \cite{Choi2012} before reconstructing the final set of OCT images. En-face images were generated for each modal data set by calculating the mean projection value along depth within a slab of 0.8 mm. Three-dimensional renderings of the BRAD-OCT volumes were generated using Fiji (3D Viewer Plugin, ImageJ 1.49) \cite{Schindelin2012}. Each modal OCT image was normalized to the total contribution by $I_{lm-norm}$ = $I_{lm}$ / ($I_{01}$ + $I_{11}$ + $I_{21}$). The BRAD ratio parameter was calculated for the higher-order modes $LP_{11}$ and $LP_{21}$ as the ratio $I_{01}$/$I_{lm-norm}$ of $LP_{01}$ to each mode. 

\subsection{Microparticles}
Polystyrene microparticles (Duke Standards, Thermo Fisher Scientific Inc.) with narrow size dispersions (mean diameter $\pm$ standard deviation) of 1.00 $\pm$ 0.01 $\mu m$, 3.00 $\pm$ 0.03 $\mu m$, 5.0 $\pm$ 0.3 $\mu m$, 7.0 $\pm$ 0.5 $\mu m$ and 10.0 $\pm$ 0.5 $\mu m$ were purchased from Distrilab BV. Aqueous suspensions with 1$\%$ particle concentration were pipetted into a 0.36 mL microwell plate and promptly imaged from the top by BRAD-OCT. Datasets covering 3.5 mm $\times$ 3.5 mm $\times$ 5.0 mm were acquired. The morphology of polystyrene beads was determined by scanning electron microscopy (FEI Quanta 200). Droplets of the aqueous microsphere suspensions were placed on a sample holder equipped with double-sided sticky carbon tape and the liquid medium was evaporated at 55$^\circ$C. Upon drying, the samples were sputter-coated with gold before SEM micrographs were recorded (acceleration voltage 15 kV).

\subsection{Brain tissue samples}
Human brain tissue samples were retrieved from the Neurobiobank of the Medical University of Vienna. A fresh brain tissue sample was obtained from a lobar resection of a gliobastoma patient (Ethics approval number EK078-2004) for investigating tumor scattering discrimination with BRAD-OCT. Two unstained, formalin-fixed samples, extracted post-mortem from two patients with Alzheimer's disease (Ethics approval number 396-2011) were used for testing the capability of BRAD-OCT for detecting highly scattering neuritic beta-amyloid plaques. A post-mortem sample extracted from the frontal cortex of a normal human brain was used as control in the studies.

\subsection{Model of light scattering by neuritic plaques}

\begin{figure*}[b]
\centering
\includegraphics[width=1\linewidth]{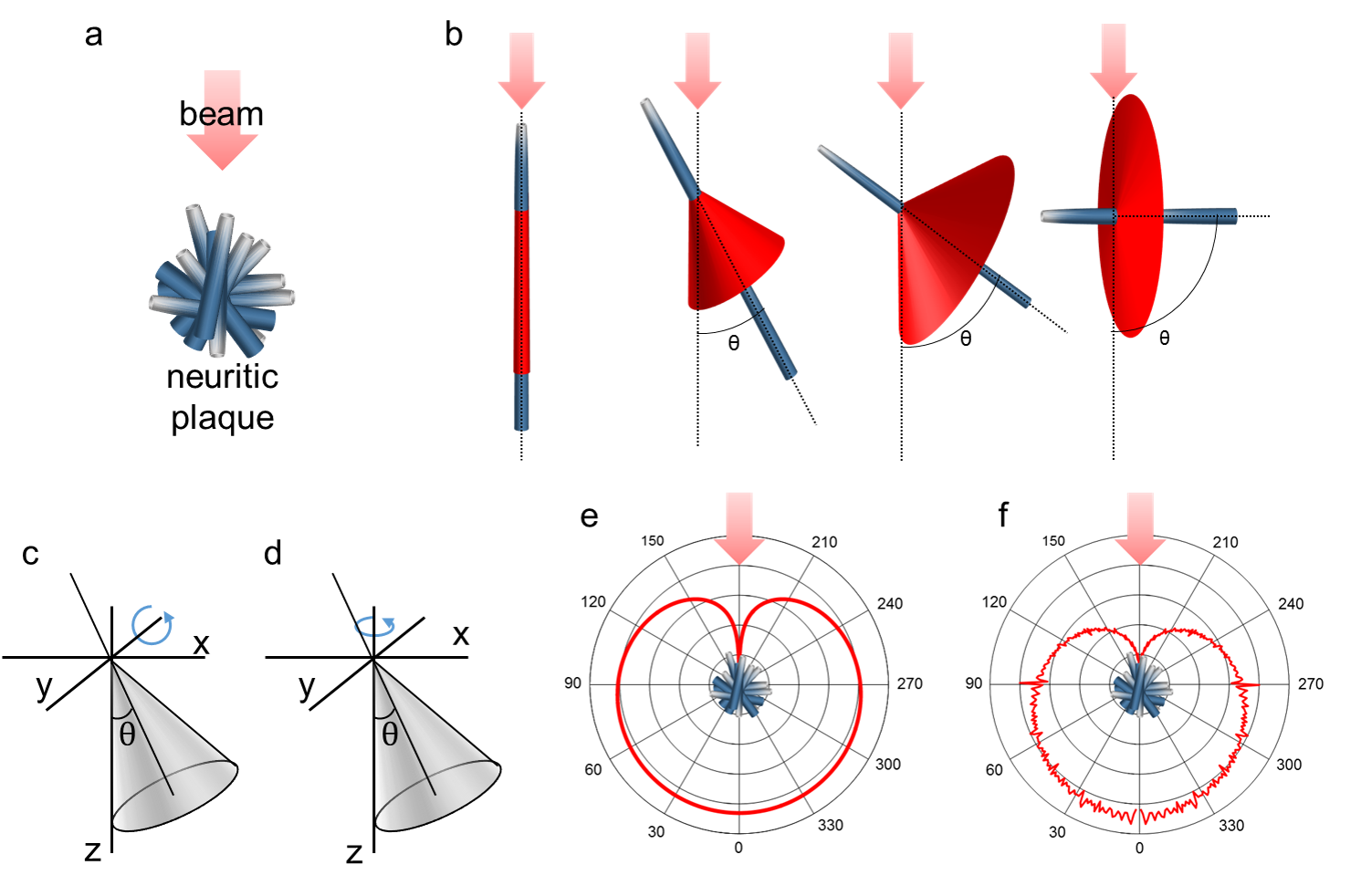}
\caption{Neuritic plaque scattering model.(a) Description of the radial distribution of the fibrils composing the neuritic plaques. (b) Scattering shape of a unique fibril oriented at different angles $\theta$ respect to the incident beam. (c) Rotation of the cone around the \textit{y}-axis. (d) Rotation around the z-axis. (e) Total scattering profile (logarithmic scale). (f) Simulated scattering profile (logarithmic scale).}
\label{fig:S2}
\end{figure*}

Numerical simulations were performed to model the angular scattering characteristics of senile amyloid-beta plaques. These neuritic plaques formed in the cerebral cortex in Alzheimer's disease consist of an accumulation of beta-amyloid protein fibrils disposed in a radiating distribution \cite{Plascencia-Villa2016,Jin2003} as shown in Fig. \ref{fig:S2}(a). The scattering profile of an individual fibril illuminated with a beam can be described with a conical shape whose angle of aperture $ \theta$ is equal to the angle between the fibril orientation and the incident beam \cite{Bohren2008}  as depicted in Fig. \ref{fig:S2}(b). We can parameterize the conical vectors radiated by a fibril as:

\begin{equation} \label{eq:12}  \left( \begin{array}{c}
S_x \\  
S_y\\
S_z
 \end{array} \right)=\left( \begin{array}{c}
z_n \tan \theta \cos v \\  
z_n \tan \theta \sin v\\
z_n
 \end{array} \right) \end{equation}
where \(0 \leq v \leq 2 \pi\) and $\theta$ is the angle between the fibril and the incident beam in a Cartesian system of coordinates where \textit{z} denotes the vertical axis and the incident beam direction. This expression describes a cone whose central axis is the \textit{z}-axis, nevertheless the cone axis must align with the orientation of the fiber itself. Thus, we introduce a rotation matrix around the \textit{y}-axis in order to tilt the cone as depicted in Fig. \ref{fig:S2}(c):

\[ \left( \begin{array}{c}
S_x \\  
S_y\\
S_z
 \end{array} \right)=\left( \begin{array}{ccc}
 \cos \theta & 0 & \sin \theta \\  
0 & 1 & 0\\
-\sin \theta & 0 & \cos \theta
 \end{array} \right) \left( \begin{array}{c}
 z_n \tan \theta \cos v \\  
z_n \tan \theta \sin v\\
z_n
 \end{array} \right)\] 
 \begin{equation} \label{eq:13}
 =
 \left( \begin{array}{c}
 z_n (1+\cos v)\sin \theta \\  
z_n \tan \theta \sin v\\
z_n(\cos \theta-\cos v \sin \theta \tan \theta)
 \end{array} \right)
\end{equation} 
For describing the fibril scattering at each azimuth orientation, another rotational matrix (this time around the \textit{z}-axis) is needed (see Fig. \ref{fig:S2}(d)):

 \[ \left( \begin{array}{c}
S_x \\  
S_y\\
S_z
 \end{array} \right)=\left( \begin{array}{ccc}
 \cos \omega & -\sin \omega & 0 \\  
\sin \omega & \cos \omega & 0\\
0 & 0 & 1
 \end{array} \right) \left( \begin{array}{c}
 z_n (1+\cos v)\sin \theta \\  
z_n \tan \theta \sin v\\
z_n(\cos \theta-\cos v \sin \theta \tan \theta)
 \end{array}\right)\]
 
\begin{equation} \label{eq:14}
 =
 \left( \begin{array}{c}
 z_n((1+\cos v)\sin \theta \cos \omega - \sin \omega \sin v \tan \theta) \\  
z_n((1+\cos v)\sin \theta \sin \omega - \cos \omega \sin v \tan \theta) \\
z_n(\cos \theta - \cos v \sin \theta \tan \theta))
 \end{array} \right)
\end{equation}

where $\theta$ is the angle between the fibril and the incident beam and $\omega$ is the angle around the vertical axis. The scattering profile of every individual fibril in the center of the plaque can be described in three dimensions as all the vectors $(S_x,S_y,S_z)^T$  from Eq. (\ref{eq:13}) that satisfy \(0 \leq v \leq 2 \pi\). Note that all fibrils produce a forward scattering component that will contribute to the final scattering pattern profile and in fact causes forward scattering to be more intense than the backward scattering. Assuming that the neuritic plaques are composed of fibrils with uniformly distributed orientations, the total scattering profile will be the sum of the contributions from all individual fibrils. Due to the rotational symmetry of the fiber arrangement assumed in our neuritic plaque, the two-dimensional scattering profile $I_{total}(\theta)$  at any orientation around the beam axis will be equal for every cross-section perpendicular to the \textit{x-y} plane. 
The two-dimensional scattering profile can be described using a sector $\theta$ of unit radius circle, $I(\theta)\varpropto \theta/2$ , as the scattering intensity for any fibril with orientation $\theta$. Hence, the total scattering in direction $\theta$ can be described as:
\begin{equation} \label{eq:15} I_{total}(\theta)\varpropto \int_{\theta}^{\pi}I(\theta ')d\theta '=\frac{\pi^2}{4}-\frac{\theta^2}{4}.\end{equation}

The total scattering profile expressed by Eq. (\ref{eq:15}) is shown in Fig. \ref{fig:S2}(e). We confirmed our model by performing a simulation based on 11,664,000 fibrils at different orientations ($90^\circ \times 360^\circ \times 360^\circ$) and adding up the scatter contributions. The results of this simulation are shown in Fig. \ref{fig:S2}(f) and agree well with the model in Eq. (\ref{eq:15}). \\

The proposed scattering model considers a uniform distribution of the fibril angles to describe neuritic plaques. While we found this rather simple model to be sufficient to roughly describe the directional scattering behavior of beta-amyloid plaques, extended features such as plaque size and a more randomized fibril distribution could be added in a more sophisticated model to achieve a more precise result. Still, our approximation substantiates the highly scattering characteristics that make plaques visible in the higher-order modes of our FMF detection.

\subsection{Histology and digital pathology}
Brain samples were formalin-fixed, paraffin-embedded and routinely processed for histopathology. Sections stained with hematoxylin and eosin were scanned with a Hamamatsu NanoZoomer 2.0 HT slide scanner. For the Alzheimer's brains, Congo red stain was used for highlighting the neuritic plaques. For the glioblastoma sample, tumor regions were annotated on the digital slide by a neuropathologist using the NDP.view2 software. The digitized slide was downsampled to 10x magnification and exported as a jpeg-image. The next processing steps were performed in Fiji \cite{Schindelin2012}. The Color Deconvolution plugin was used to obtain an 8-bit gray-scale image of the hematoxylin stain \cite{Ruifrok2001}. Phansalkar automated local thresholding was performed to segment the individual cell nuclei \cite{Phansalkar2011}. Small aberrantly stained fragments were discarded by perfoming the binary Open operation. Densely clustered nuclei were separated with the inbuilt watershed function. The thereby obtained cell mask was loaded into MATLAB R2014b (MathWorks) and analyzed in blocks of 160 $\times$ 160 pixels ($\sim$146 $\times$ 146 $\mu m$). For each block, the count and mean size of nuclei were calculated. Nuclear density and nuclear size data from all blocks were visualized as heat maps.

\section{Results}
\subsection{Angular response of fiber modes}

In our experimental realization using a short-cavity laser sweeping from 1240 nm to 1380 nm wavelengths at 100 kHz and a Mach Zehnder interferometer setup based on standard SMF-28e+ telecom single-mode fibers, we implemented a step index FMF with 25-$\mu m$ core size and an NA of 0.10 for the detection. A polarization-based circulator directed light emitted by the SMF onto the sample and rerouted backscattered light to the FMF (Fig. \ref{fig:fig1}(a)). Light coupled into and guided by the FMF ultimately interferes with the reference light and is detected at up to 350 MHz. 	\par
\begin{figure}[htb]
\centering
\includegraphics[width=0.7\linewidth]{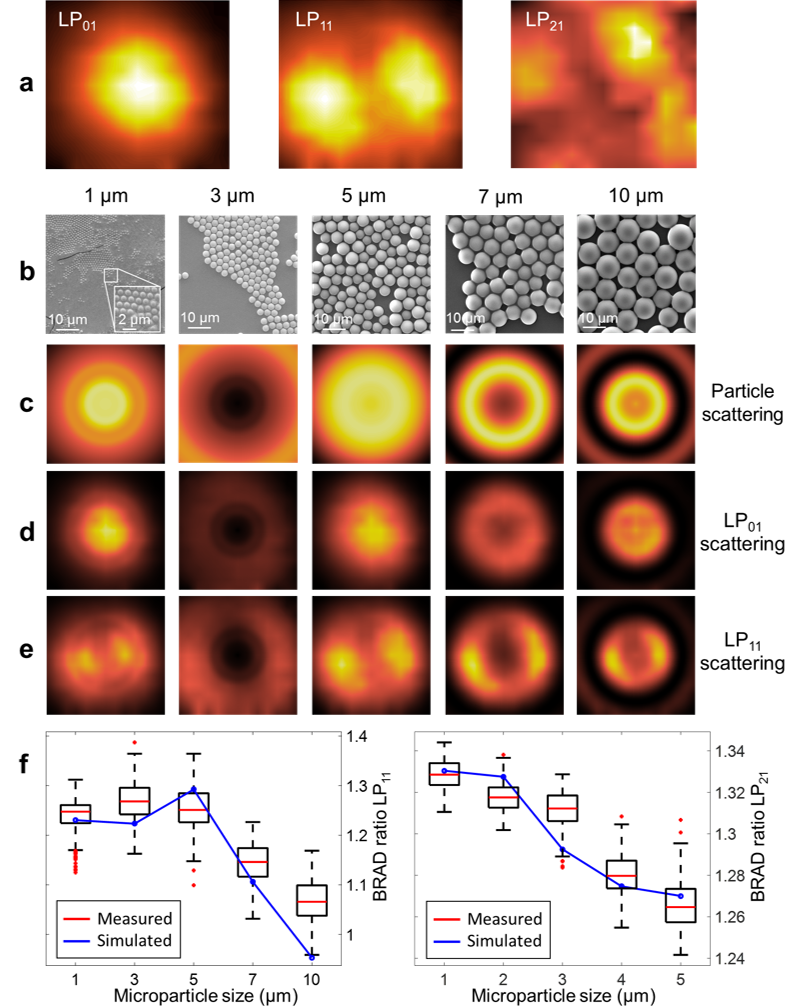}
\caption{Angular response of fiber modes and BRAD ratios for different particle sizes. (a) Angular FMF response to different angle orientations in a field of view of $\pm 4^\circ$ of a mirror in the sample position. The fiber angular transfer function shows good agreement with the modal intensity profiles simulated in Fig. \ref{fig:S1}. (b) Electron microscopy scans of the different particle sizes. (c) Angular scattering profiles were simulated for different microparticle sizes using Mie theory. The light scattered from the polymer microparticles excites the FMF modes according to the angular response measured for the FMF as shown in (d) and (e) for $LP_{01}$ and $LP_{11}$. (f) Results of the BRAD ratio between $LP_{01}$/$LP_{11}$ and $LP_{01}$/$LP_{21}$ for the different microparticle sizes measured with BRAD-OCT and the corresponding theoretical results displayed in red.}
\label{fig:fig2}
\end{figure}

In order to investigate the response of the FMF to sample light impinging at different angles, we placed a mirror into the sample position and tilted it at equidistant angular steps covering the entire detection field of view of the FMF of 8$^\circ$. Signals corresponding to the intensity of light coupled into the modes were mapped into different positions within the depth range. The measured angular fiber transfer function is shown in Fig. \ref{fig:fig2}(a) for the three most prominent modes ($LP_{01}$, $LP_{11}$ and $LP_{21}$). The fundamental mode ($LP_{01}$) reaches its maximum for directly backreflected light ($\vartheta =0^\circ$) whereas the higher-order $LP_{11}$ and $LP_{21}$ modes have a minimum there and unveil their maximum response at greater angles of reflection. Clearly, the angular response of the modes emulates the modal intensity patterns propagating from the fiber tip (Fig. \ref{fig:S1}). The interrelationship of angle of reflection and intensity distribution allotted to the fiber modes led us to hypothesize that, reversely, angular scattering characteristics of tissues may be deduced from the modal intensity partitioning observed in OCT scans.

\subsection{BRAD-OCT differentiates Mie scatterers based on their angular scattering}
The intensity and angular characteristic of scattered light depends on material parameters such as scatterer shape, size and refractive index \cite{Jacques2013,Bohren2008}. For scattering particles with spherical shape sized similar to the wavelength of the interrogation light, the scattering behavior is described by Mie scattering \cite{Jacques2013,Bohren2008}. Different polystyrene spheres with diameters ranging from 1 $\mu m$ to 10 $\mu m$ in an aqueous suspension with a concentration of 1$\%$ solids (Fig. \ref{fig:fig2}(b)) were used for testing the BRAD-OCT scattering detection. The two-dimensional angular scattering profiles were calculated for the different microspheres (Fig. \ref{fig:fig2}(c)). Distinct circular scattering patterns can be observed which greatly differ for the diverse particle sizes. Based on the modal angular response measured for the FMF detection of our system (Fig. \ref{fig:fig2}(a)), we simulated the modal excitation patterns for light scattered from the differently sized particles (see Fig. \ref{fig:fig2}(d) for $LP_{01}$ and Fig. \ref{fig:fig2}(e) for $LP_{11}$). Aiming for an angular scattering parameter, we computed the ratio of the total intensity picked up by the fundamental mode $LP_{01}$ mode over the light intensity launched into $LP_{11}$ (see Methods). As $LP_{01}$ favorably collects directly backscattered light (i.e. bright field OCT detection) and the higher modes predominantly pick up light scattered at greater angles (i.e. dark field OCT detection), we termed this relationship the BRAD ratio (see Methods for additional details). The dependence of the BRAD ratio on particle size is plotted in Fig. \ref{fig:fig2}(f). \par
Next, we validated the theoretically expected BRAD ratios by imaging aqueous suspensions of polystyrene microspheres (particle diameters: 1.0 $\mu m$, 3.0 $\mu m$, 5.0 $\mu m$, 7.0 $\mu m$ and 10.0 $\mu m$). The acquired OCT images were demultiplexed to extract the modal intensity contributions for $LP_{01}$, $LP_{11}$ and $LP_{21}$ (see Methods). The BRAD ratios were computed from the average intensities in sample volumes of 125 mm$^3$ and were in good agreement with the theoretical estimates (see Fig. \ref{fig:fig2}(f)). These measurements showcase the potential of BRAD for OCT imaging with contrast generated by angular scattering related to particle size.

\begin{figure}[tbp]
\centering
\includegraphics[width=0.7\linewidth]{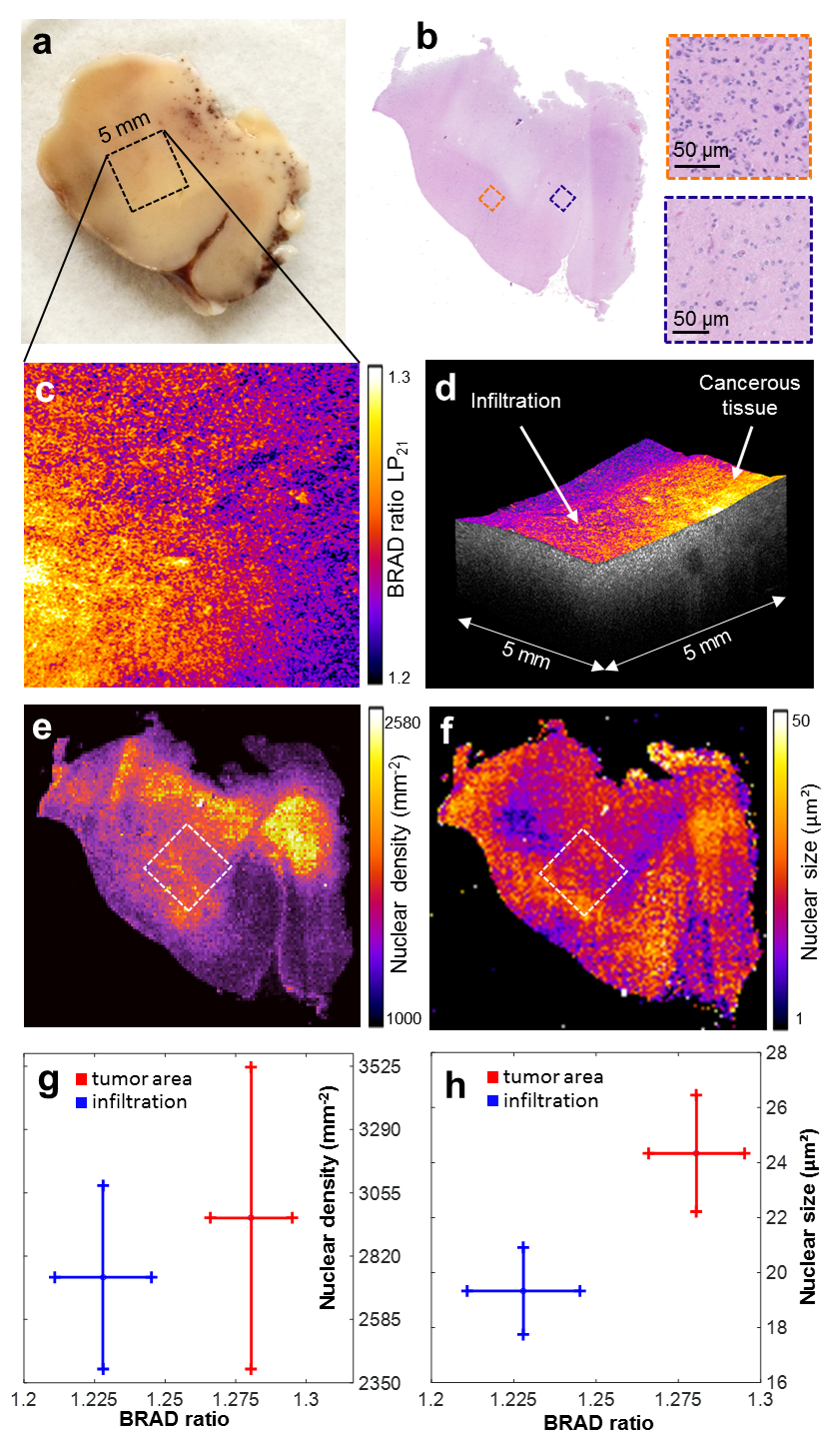}
\caption{BRAD-OCT in cancer imaging. (a) Macroscopic image of a temporal lobe resection partly infiltrated by glioblastoma (note petechial bleedings in the white matter). The scanned area is indicated by a square. (b) En face map displaying the BRAD ratio of $LP_{21}$ and $LP_{01}$ revealed different scattering characteristics for the cancerous area as compared to the adjacent brain parenchyma. (c) The corresponding histological scan of the brain sample after OCT imaging revealed adjacent brain parenchyma of low cellularity (blue box). In contrast, tumorous tissue (orange box) is associated with increased cell density and wide variations in the nuclear size and shape. (d) Representative OCT volume of the scanned area with overlaid BRAD map indicating the tumor area and infiltration zone. (e) Nuclear density map by digital histology. (f) Nuclear size map. (g) Comparison of BRAD ratio with nuclear density in an area of tumorous and adjacent tissue, respectively. (h) Comparison of BRAD ratio with nuclear size.}
\label{fig:fig3}
\end{figure}

\subsection{BRAD-OCT discriminates tumor in glioblastoma}

Cancerous lesions are characterized by abnormal cell growth including changes of both the cellular ultrastructure and microstructure. Altered tissue morphology is, in many cases, directly linked to changes of the optical tissue properties. Optical sensing and imaging approaches therefore provide a rapid and non-invasive means for the assessment of cancerous lesions \cite{Hillman2007}. Here we apply BRAD-OCT for imaging ex vivo native tissue of a human glioblastoma patient (Fig. \ref{fig:fig3}(a)). Glioblastoma is the most frequent malignant brain tumor representing 49 $\%$ of all primary malignant brain tumors and the most malignant histologic type (WHO grade IV) \cite{Ostrom2016}. With fewer than 6$\%$ of glioblastoma patients still alive at five years after diagnosis, the prognosis is extremely poor \cite{Ostrom2016}. On microscopic evaluation, tumor tissue is characterized by an increased cell density and highly variable cell morphology compared to healthy brain parenchyma (Fig. \ref{fig:fig3}(b)). Furthermore, glioblastoma typically shows a diffusely infiltrating growth pattern, thereby forming a poorly delineated border zone between cellular tumor and the surrounding CNS tissue.\par

BRAD-OCT images of the infiltration zone of the tumor invading into the adjacent brain parenchyma were acquired. We calculated the BRAD ratio for the higher-order modes with respect to the fundamental mode $LP_{01}$ to identify changes in the tissue scattering pattern (see Methods). The calculated BRAD ratio for $LP_{21}$ shows scattering differences between compact tumor areas and adjacent brain tissue (Fig. \ref{fig:fig3}(c)). A volume rendering fused with the BRAD ratio information provides a combined overview of the 3D picture and tissue scattering information (Fig. \ref{fig:fig3}(d)). In order to calibrate and validate our BRAD image data to the gold standard, we performed a histopathologic examination of the tissue sample and quantitatively analyzed cellularity and cellular morphology (see Methods). Unsurprisingly, we observed relatively high nucleus density in the compact tumor (Fig. \ref{fig:fig3}(e)) and a strong variation of the nuclear sizes (Fig. \ref{fig:fig3}(f)). Moreover, the histological analysis (Fig. \ref{fig:fig3}(e-f)) revealed substantial intratumoral heterogeneity and widespread areas of diffuse infiltration within the sample. BRAD-OCT images notably revealed a difference in the scattering signature within tumor-affected areas while the healthy tissue was characterized by a low BRAD ratio. In order to obtain quantitative correlations of our data with the histology results, we analyzed two areas of 0.6 mm $\times$ 0.6 mm by computing the BRAD ratio in tumor region and adjacent brain parenchyma independently. The BRAD ratio of the scanned areas was then directly compared to the number of nuclei (Fig. \ref{fig:fig3}(g)) and the nuclear size (Fig. \ref{fig:fig3}(h)). While the mean values of the nuclear density varied for different BRAD ratios, the corresponding standard deviations were too large to allow unequivocal tissue differentiation. In the tumor area, the average nuclear size (24.3 $\pm$ 2.0 $\mu m^2$) was significantly larger (p = 0.0046) and the BRAD ratio (1.280 $\pm$ 0.017) was significantly higher than in the infiltration zone (nuclei: 19.4 $\pm$ 1.5 $\mu m^2$, BRAD ratio = 1.220 $\pm$ 0.020). However despite a substantial variation of the nuclear density in both areas, the difference between the tumor area (2961 $\pm$ 526 mm$^{-2}$) and the adjacent brain parenchyma (2740  $\pm$ 320 mm$^{-2}$) was not statistically significant (p = 0.167).

\subsection{BRAD contrasts neuritic beta-amyloid plaques in Alzheimer's disease}
\begin{figure*}[htbp]
\centering
\includegraphics[width=1.02\linewidth]{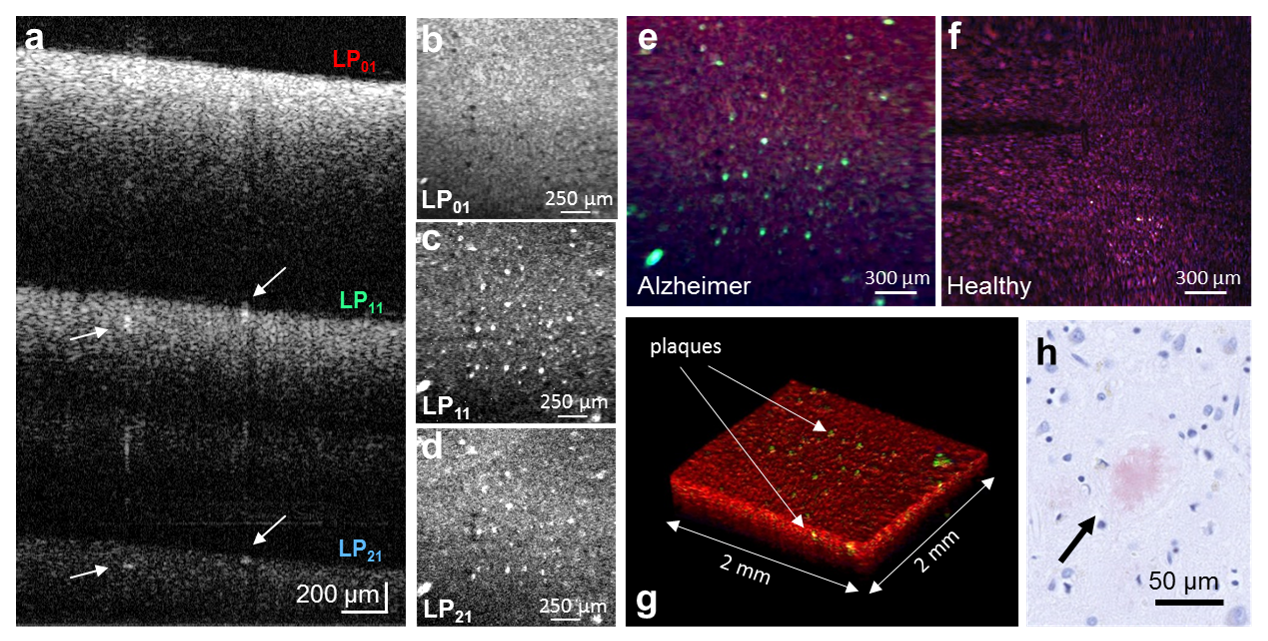}
\caption{BRAD-OCT of neuritic plaques in Alzheimer's disease. (a)  OCT depth-multiplexed B-scan images of a brain from a late-stage Alzheimer's patient for the $LP_{01}$, $LP_{11}$ and $LP_{21}$ modes. Neuritic plaques can be observed as hyper-scattering structures particularly showing up in the B-scans obtained by the higher-order modes. (b-d) En-face intensity projection images for $LP_{01}$, $LP_{11}$, and $LP_{21}$. The bright field scattering collected by the fundamental mode $LP_{01}$ (b) does not enable clear differentiation of neuritic plaques while the higher modes (c - d) enhance the contrast between plaques (arrows in (a)) and the surrounding tissue. The combination of the modal en face projections in the channels of an RGB image provides an easy discrimination between (e) Alzheimer affected and (f) non-affected tissue, highlighting the neuritic plaques in the green channel. (g) Volume rendering encoding all modal information in an RGB channel. (h) Histologic section of a neuritic plaque highlighted with Congo red stain scanned after applying BRAD-OCT.}
\label{fig:fig4}
\end{figure*}

Owing to the angular detection differences of FMF modes, BRAD provides OCT with contrast for the directionality of light scattering. We next demonstrate BRAD for OCT imaging of cerebral pathology with pronounced directional scattering characteristics. Neuritic plaques are one hallmark of Alzheimer's disease and are composed of beta-amyloid fibrils arranged in a radiating geometry \cite{Plascencia-Villa2016,Baumann2017}. Unlike the surrounding gray matter tissue with its rather pronounced backward and forward scattering characteristics \cite{Lee2016}, the fibrous structure of neuritic plaques produces a distinct angular scattering pattern (Fig. \ref{fig:S2}(f)). Post-mortem formalin-fixed cortical brain tissue from a human patient with late-stage Alzheimer's disease was imaged using BRAD-OCT. In the NIR regime used by our prototype, the lion's share of the plaque load cannot be sufficiently contrasted on the fundamental mode $LP_{01}$ as indicated in the B-scan image of Fig. \ref{fig:fig4}(a) whereas the plaques are highlighted as hyper-intense structures in the higher-order modes $LP_{11}$ and $LP_{21}$. The en face projection of the bright field OCT image constructed by $LP_{01}$ (Fig. \ref{fig:fig4}(b)) cannot portray the neuritic plaques while the dark field OCT images enabled by the higher-order modes provide enhanced visualization of the neuritic plaques (Fig. \ref{fig:fig4}(c-d)). By combining bright and dark field contrast in an RGB image where the $LP_{01}$, $LP_{11}$ and $LP_{21}$ signals are fed into the red, green and blue channels, respectively, the neuritic plaques can easily be localized in surrounding gray matter tissue (see projection images in Fig. \ref{fig:fig4}(e-f) and volume rendering in Fig. \ref{fig:fig4}(g)). Histology was performed after imaging to confirm the corresponding hyperscattering neuritic plaques (Fig. \ref{fig:fig4}(h)). 

\section{Discussion}
By combining the advantages of OCT imaging with the transmission properties of few-mode fibers, we have demonstrated a new concept for multiplexed biomedical imaging with simultaneous bright and dark field detection. The beauty of our approach lies in its simplicity. Using an off-the-shelf FMF (commercially available for \texteuro 25 per meter), we gained two additional detection channels and, therewith, sensitivity to the directionality of scattering. We investigated the impact of particle size on the measured BRAD signal and validated the size-signal relation in microsphere phantoms. The direct connection between scattering and biological composition enables the assessment of nuclear morphology via optical characterization \cite{Chalut2008,Yang2012,Kalashnikov2012}. As a first demonstration for biomedical imaging, we then showed the potential of BRAD by contrasting lesions related to brain cancer and Alzheimer's disease. On top of the intrinsic scatter contrast exploited here, sensitivity for specific structures and compounds may be gained by using BRAD with dedicated extrinsic contrast agents such as gold nanoparticles \cite{Jans2009,Liba2016,Lippok2017}. \par
In this work, we used a step index FMF with a fiber parameter of $V=9$ supporting only a small number of modes. By using a fiber with a greater $V$ parameter - for instance by using a fiber with greater core size or a greater profile height parameter, or by switching to a light source operating in a shorter wavelength regime - the number of modes could be increased at the cost of less average light intensity per mode owing to the spread of the available light intensity over more modes. While we found the low number of modes in our implementation appropriate for BRAD-OCT imaging because of imaging depth constrictions, a greater number of modes may enable a more detailed characterization of the angular scattering profile and hence a profound estimation of the size parameters of scattering particles, which thus far has only been possible using sophisticated scanning and acquisition schemes \cite{Desjardins2007,Zhu2011,Wartak2017}. \par
The main drawback of the proposed method is the sensitivity drop in the higher modes due to modal transmission efficiency, as can be observed in the low intensity of the $LP_{21}$ images. This is due to the fact that the main part of the light coupled into the fiber is collected by $LP_{01}$ and that the polarization of the reference arm was set to maximize the sensitivity of this given mode and not fully matching the polarization state with the higher modes. Having a second detection channel for the cross-polarized state might provide information extra information and recover polarization induced losses in the high modes.\par
While we used a table top OCT microscope assembly for imaging ex vivo samples to prove the concept of FMF-powered BRAD in this work, fiber-optic setups do enable flexible arrangements in scan heads, surgical microscopes, and endoscopic probes \cite{Adler2007,Tearney1996}. Also, the fields of application of BRAD-OCT are not limited to neuroimaging. FMF detection for OCT and BRAD could be implemented in virtually any field OCT has proven useful for, ranging from ophthalmology and cardiology to non-destructive testing in materials science. In summary, we have presented a new detection concept for optical systems which exploits the capabilities of FMFs providing enhanced contrast and demonstrated its potential for medical applications such as neuroimaging.

\section*{Funding}
European Research Council (ERC StG 640396 OPTIMALZ); \\Austrian
Science Fund (FWF grant P25823-B24; FWF grant KLI 394; FWF grant P26553-N20);\\

\section*{Acknowledgments}
The authors thank Simon Strehn for fiber testing and Andreas Hodul for technical assistance. We would also like to acknowledge Stanislava Fialov\'{a} for her support setting up the BRAD-OCT system. We wish to thank Laurin Ginner, Florian Beer, Richard Haindl, Rainer Leitgeb and Gabor G. Kovacs for their assistance and helpful discussions.

\section*{Disclosures}
The authors declare that there are no conflicts of interest related to this article.

\end{document}